\documentclass{article}
\usepackage{spconf,amsmath,graphicx}
\usepackage{mathrsfs}
\usepackage{float}
\usepackage{amssymb}
\usepackage{cite}
\usepackage{booktabs}
\usepackage{multicol,listings,multirow}
\usepackage{lipsum}
\usepackage{tikz}
\usepackage{amssymb}
\usepackage{wrapfig}
\usepackage{array}
\usepackage{makecell}
\usepackage{booktabs}
\usepackage{array}
\usepackage{color}
\usepackage{makecell}
\usepackage[colorlinks=true,
            linkcolor=blue,
            filecolor=blue,      
            urlcolor=blue,       
            citecolor=blue
            ]{hyperref}
\usepackage[marginal]{footmisc}
\usepackage[misc]{ifsym}
\usepackage{soul}
\usepackage{enumitem}
\setlist{nosep, leftmargin=14pt}

\usepackage{mwe} 

\title{Unsupervised Domain Adaptation for Cross-Modality Retinal Vessel Segmentation via Disentangling Representation Style Transfer and Collaborative Consistency Learning}
\name{Linkai Peng$^{1\ddagger}$, Li Lin$^{1,2\ddagger}$, Pujin Cheng$^{1}$, Ziqi Huang$^{1}$, Xiaoying Tang$^{1*}$
\thanks{$^\ddagger$ {These authors contributed equally}.}}
\address{$^1$Department of Electronic and Electrical Engineering, Southern University of Science and Technology,\\ Shenzhen, China\\
$^2$Department of Electrical and Electronic Engineering, The University of Hong Kong,\\Hong Kong SAR, China}
\begin{document}
%
\maketitle
\begin{abstract}
Various deep learning models have been developed to segment anatomical structures from medical images, but they typically have poor performance when tested on another target domain with different data distribution. Recently, unsupervised domain adaptation methods have been proposed to alleviate this so-called domain shift issue, but most of them are designed for scenarios with relatively small domain shifts and are likely to fail when encountering a large domain gap. In this paper, we propose DCDA, a novel cross-modality unsupervised domain adaptation framework for tasks with large domain shifts, e.g., segmenting retinal vessels from OCTA and OCT images. DCDA mainly consists of a disentangling representation style transfer (DRST) module and a collaborative consistency learning (CCL) module. DRST decomposes images into content components and style codes and performs style transfer and image reconstruction. CCL contains two segmentation models, one for source domain and the other for target domain. The two models use labeled data (together with the corresponding transferred images) for supervised learning and perform collaborative consistency learning on unlabeled data. Each model focuses on the corresponding single domain and aims to yield an expertized domain-specific segmentation model. Through extensive experiments on retinal vessel segmentation, our framework achieves Dice scores close to target-trained oracle both from OCTA to OCT and from OCT to OCTA, significantly outperforming other state-of-the-art methods.

\end{abstract}
\section{Introduction}
\label{sec:intro}
\vspace{-0.2cm}
\begin{figure}[htbp]
	\centering
	\vspace{-0.2cm}
	\setlength{\abovecaptionskip}{-0.2cm}   
	\setlength{\belowcaptionskip}{-1cm}   
	\centerline{\includegraphics[height=2.6cm]{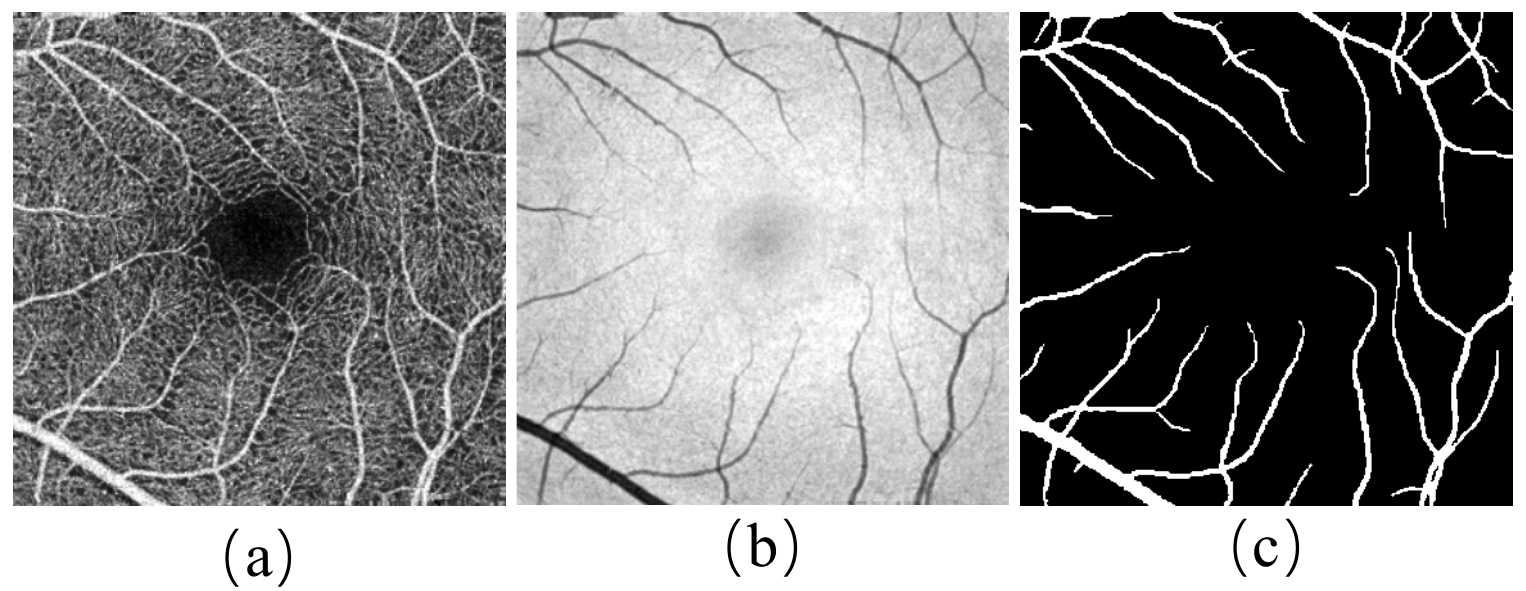}}
	\vspace{-0.3cm}
	\caption{(a) is an OCTA image, (b) is an OCT image, and (c) is the corresponding RV segmentation label.}\medskip
	\vspace{-0.7cm}
	\label{intro}
\end{figure}
Optical coherence tomography (OCT) and optical coherence tomography angiography (OCTA) are two novel and non-invasive ophthalmic imaging modalities. OCT has micrometer-level resolutions both longitudinally and laterally \cite{huang1991optical} and OCTA can generate volumetric angiography images in a fast and detailed fashion \cite{de2015review}. These two imaging modalities can both produce detailed visualizations of vascular structures within the retina, exhibiting great potential for improving the diagnosis accuracy of eye-related diseases \cite{osareh2002classification, lin2021bsda}. Examples of these two modalities are shown in Figure \ref{intro}. For segmenting retinal vessels (RVs) from OCT and OCTA images for specific clinical purposes, many deep learning methods have been developed \cite{islam2020deep, peng2021fargo}. Both 2D and 3D volumes have been utilized to explore various RV segmentation approaches.

In clinical practice, RV segmentation labels are typically annotated on OCTA images due to its relatively clearer visualization of vascular structures. However, segmentation models trained on OCTA images usually perform poorly when tested on OCT images because of domain shift \cite{saenko2010adapting}. Thus, it is clinically meaningful to design a RV segmentation framework trained on OCTA images and corresponding labels, but can also work well for segmenting OCT images. In order to solve domain shift related issues, a number of domain adaptation methods have been developed, among which unsupervised domain adaptation (UDA) methods have been gaining greatest popularity in recent years. For instance, Zhou et al. \cite{Zou_2018_ECCV} created a segmentation workflow via class-balanced self-training. Tzeng et al. \cite{tzeng2017adversarial} introduced adversarial training into their UDA framework. Further, Tuan-Hung et al. \cite{vu2019advent} utilized the idea of entropy minimization to design an effective UDA method and Yi-Hsuan et al. \cite{tsai2018learning} improved it by adopting a multi-level adversarial learning scheme. Judy et al. \cite{hoffman2018cycada} employed a cycle-consistent adversarial network (CycleGAN). These methods nevertheless contain only one segmentation model, which is one of the main reasons why they fail to converge when the domain gap between training and testing is considerably large. Meanwhile, CycleGAN assumes a one-to-one mapping from source domain to target domain, and thus lacks diversity in the image translation outputs. Moreover, these aforementioned methods are typically designed for addressing small domain gaps, e.g., from GTA5 \cite{richter2016playing} to Cityscapes \cite{cordts2016cityscapes}. Their performance will decrease remarkably when the data distributions of the source domain and the target domain are quite different, as we will show later.

In such context, we propose a novel UDA framework combining a disentangling representation style transfer (DRS\\T) module and a collaborative consistency learning (CCL) module, which alleviates the aforementioned issues and exhibits superior performance. For DRST, we use a disentangled representation framework under the assumption that images can be decomposed into content components and style codes \cite{lee2020drit++}. Multiple tasks with different objective functions are applied to guide the network to separate apart domain invariants (content components) and domain variants (style codes). Adversarial training is used to train the network to produce images with contents from one domain and styles from the other domain to fool the discriminator. Cycle-consistency is also employed for image reconstruction, which encourages the reconstructed images to be close to the original images. For CCL, we have a source-style segmentation model $\mathcal{F^{S}}$ and a target-style segmentation model $\mathcal{F^{T}}$. We first train $\mathcal{F^{S}}$ on source domain with labeled data, which can be viewed as an expert on the source domain. Then $\mathcal{F^{S}}$ and $\mathcal{F^{T}}$ are trained with both original images and transferred images generated by DRST. A collaborative consistency loss is utilized to minimize the distribution differences between the corresponding outputs of the two models.

The main contributions of this paper are three-fold: (1) We propose an innovative UDA framework for RV segmentation across OCT and OCTA images. (2) The UDA framework builds its basis on disentangling representation style transfer and collaborative consistency learning, which can be effectively applied to settings with large domain gaps. We validate our framework on RV segmentation from OCT and OCTA images. (3) Extensive comparison experiments are conducted, both quantitatively and qualitatively, showing that our proposed pipeline achieves superior performance over representative state-of-the-art (SOTA) methods on the publicly-available OCTA-500 dataset \cite{li2020image}. We make our source code available at \url{https://github.com/lkpengcs/DCDA}.

\vspace{-0.3cm}
\section{Method}
\label{sec:method}
\begin{figure*}[htbp]
	\centering
	\vspace{-1.7cm}
	\setlength{\abovecaptionskip}{-0.2cm}   
	\setlength{\belowcaptionskip}{-1cm}   
	\centerline{\includegraphics[height=8.7cm]{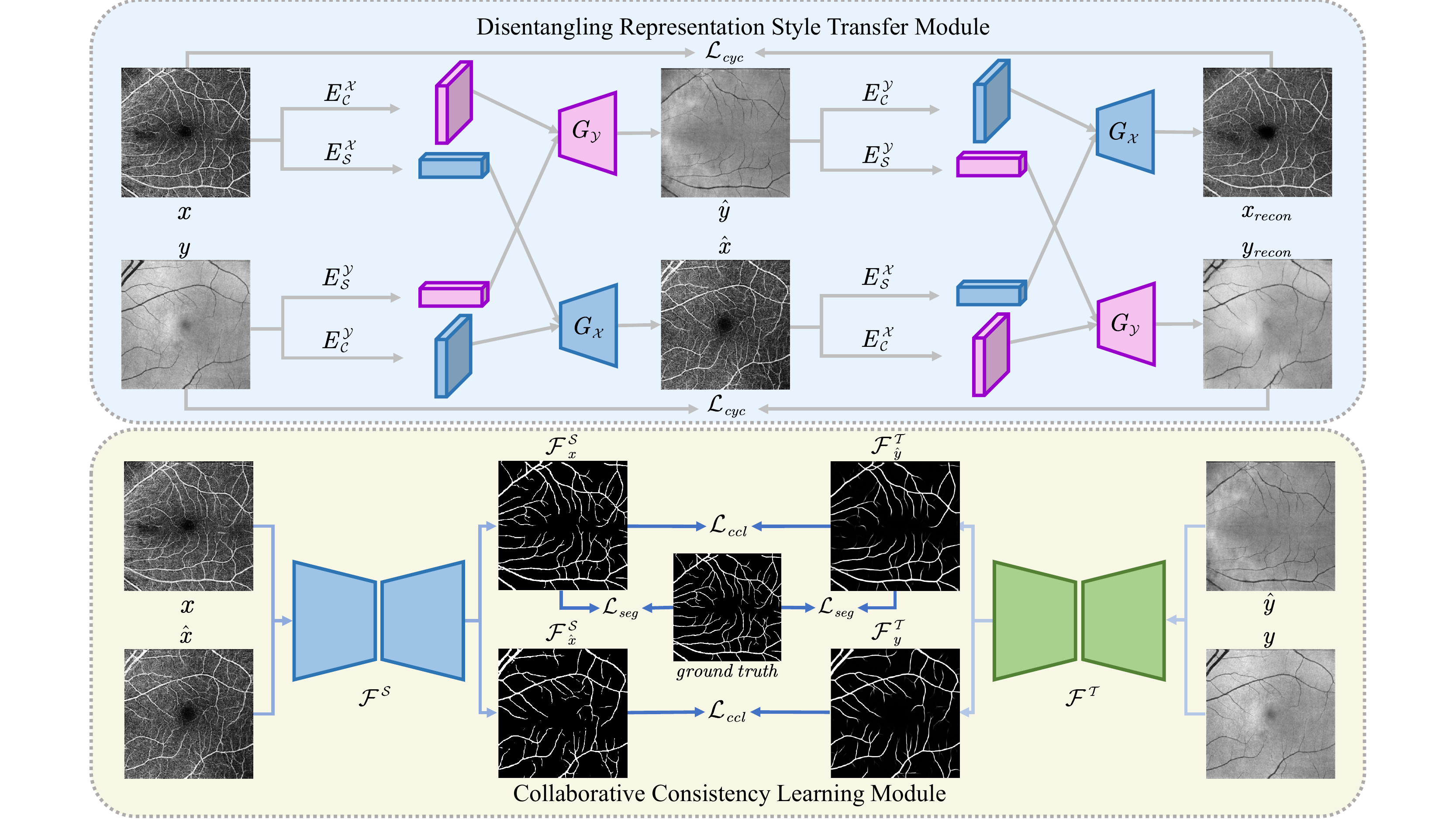}}
	\vspace{-0.3cm}
	\caption{Schematic demonstration of the architecture of our DCDA framework. The upper part represents the disentangling representation style transfer module and the lower part represents the collaborative consistency learning module.}\medskip
	\vspace{-0.5cm}
	\label{model}
\end{figure*}

The proposed framework is shown in Figure \ref{model}, which consists of a disentangling representation style transfer module and a collaborative consistency learning module.

\subsection{Disentangling Representation Style Transfer}
\label{subsec:i2i}

\subsubsection{Representation Disentanglement}
\label{subsubsec:disentangle}

Let $x \in \mathcal{X}, y \in \mathcal{Y}$ respectively be unpaired images from a source domain and a target domain. To separate apart content and style features, we use content encoders $E\mathcal{_{C}^{X}}$ and $E\mathcal{_{C}^{Y}}$ to extract content features $\mathcal{C_{X}}$ and $\mathcal{C_{Y}}$ from images in the source and target domains. We share weight between the last layer of the two content encoders due to the assumption that images should share a same content latent space. Style codes $\mathcal{S_{X}}$ and $\mathcal{S_{Y}}$ can be obtained similarly by style encoders $E\mathcal{_{S}^{X}}$ and $E\mathcal{_{S}^{Y}}$ without any weight-sharing strategy since style codes are assumed to be domain-specific. We also use a content discriminator $\mathcal{D_{C}}$ to discriminate the extracted content features in the two different domains, enforcing $E\mathcal{_{{C}}^{X}}$ and $E\mathcal{_{{C}}^{Y}}$ to produce encoded content features in a common latent space.

\vspace{-0.2cm}
\subsubsection{Adversarial Training and Cycle Consistency}
\label{subsubsec:advcyc}

Our pipeline also contains two generators, $G_\mathcal{X}$ and $G_\mathcal{Y}$. In the first translation step, $G_\mathcal{X}$ uses $\mathcal{S_{X}}$ and $\mathcal{C_{Y}}$ to generate source-style images $\hat{x}$ and $G_\mathcal{Y}$ uses $\mathcal{S_{Y}}$ and $\mathcal{C_{X}}$ to generate target-style images $\hat{y}$. We apply the adversarial loss \cite{goodfellow2014generative} to both translation directions. Discriminators $\mathcal{D_{X}}$ and $\mathcal{D_{Y}}$ are used to discriminate between real images and fake images generated by $G_{\mathcal{X}}$ and $G_{\mathcal{Y}}$. Taking source domain as an example, the loss function can be expressed as below
\vspace{-0.2cm}
\begin{equation}
\begin{aligned}
\label{eq1}
\mathcal{L}_{adv} &= \mathbb{E}_{\boldsymbol{x} \sim p_{\text {x }}(\boldsymbol{x})}[\log \mathcal{D_{X}}(\boldsymbol{x})]\\&+  \mathbb{E}_{\boldsymbol{y} \sim p_{\boldsymbol{y}}(\boldsymbol{y})}[\log (1-\mathcal{D_{X}}(\boldsymbol{\hat{x}}))].
\end{aligned}
\end{equation}

In the second translation step, we use $\hat{x}$ and $\hat{y}$ as inputs for $E\mathcal{_{C}^{X}}$ and $E\mathcal{_{C}^{Y}}$. By swapping the style codes again, the two generators $G_\mathcal{{X}}$ and $G_\mathcal{{Y}}$ are expected to reconstruct the original images. A cycle-consistency loss \cite{zhu2017unpaired} is applied to enforce such constraint, namely
\vspace{-0.2cm}
\begin{equation}
\begin{aligned}
\label{eq2}
\mathcal{L}_{\text {cyc }} &=\mathbb{E}_{x \sim p_{\text {x }}(x)}\left[\|G\mathcal{_{X}}(E\mathcal{_{S}^{X}}(\hat{x}), E\mathcal{_{C}^{Y}}(\hat{y}))-x\|_{1}\right] \\
&+\mathbb{E}_{y \sim p_{\text {y }}(y)}\left[\|G\mathcal{_{Y}}(E\mathcal{_{S}^{Y}}(\hat{y}), E\mathcal{_{C}^{X}}(\hat{x}))-y\|_{1}\right].
\end{aligned}
\end{equation}

\subsection{Collaborative Consistency Learning}
\label{subsec:segmodel}

Traditional UDA methods use only one segmentation model trained on both source domain and target domain and are likely to converge to non-optimal values when the domain gap is large. Thus, we design a collaborative consistency learning module containing two segmentation models; each model is expected to be an expert on each single domain. We first train our source-style segmentation model $\mathcal{F^{S}}$ on source domain with the supervision of $ground$ $truth$ to get a source-style expert. Then $x$ and $\hat{x}$ are fed into $\mathcal{F^{S}}$ to get outputs $\mathcal{F^{S}}_{x}$ and $\mathcal{F^{S}}_{\hat{x}}$. We input $y$ and $\hat{y}$ into $\mathcal{F^{T}}$ and obtain $\mathcal{F^{T}}_{y}$ and $\mathcal{F^{T}}_{\hat{y}}$. Typically, $\mathcal{F^{T}}_{\hat{y}}$ is supervised by $\mathcal{F^{S}}_{x}$ and $ground$ $truth$, while $\mathcal{F^{T}}_{y}$ is encouraged to be close to $\mathcal{F^{S}}_{\hat{x}}$.

During training, we set a starting flag $\tau$ when the source-style segmentation model $\mathcal{F^{S}}$ converges to a relatively optimal point, before which only a segmentation loss $\mathcal{L}_{seg}$ is applied. Then we train $\mathcal{F^{S}}$ and $\mathcal{F^{T}}$ together and a  collaborative consistency loss $\mathcal{L}_{ccl}$ is added to minimize the distribution differences between the corresponding outputs of $\mathcal{F^{S}}$ and $\mathcal{F^{T}}$. Available $ground$ $truth$ is also used for supervising $\mathcal{F^{T}}$ via adding another segmentation loss $\mathcal{L}_{seg}$ between $\mathcal{F^{T}}_{\hat{y}}$ and $ground$ $truth$. The total loss of CCL is formulated as in Eq. \ref{eq3}. For these two models, we use ResNet \cite{he2016deep} as the encoder of our U-net \cite{ronneberger2015u} architecture to achieve relatively superior performance on RV segmentation.
\vspace{-0.1cm}
\begin{equation}
\label{eq3}
\mathcal{L}_{\text {joint }}=\left\{\begin{array}{cl}
\mathcal{L}_{\text {seg }}, & \text { epoch } \leqslant \tau, \\
\mathcal{L}_{\text {seg }}+\mathcal{L}_{\text{ccl}}, & \text { epoch }>\tau.
\end{array}\right.
\vspace{-0.2cm}
\end{equation}

\vspace{0.1 cm}
\section{Experiments}
\label{sec:experiment}
\vspace{-0.2cm}
\subsection{Datasets and Image Preprocessing}
\label{sec:data}
\vspace{-0.1cm}
The OCTA-500 dataset is divided into two subsets according to field of view (FOV). One subset contains 300 samples with $6mm \times 6mm$ FOV, named 6M. The other subset contains 200 samples with $3mm \times 3mm$ FOV, named 3M \cite{li2020image}. We utilize \emph{en-face} OCT images generated by average projection between outer plexiform layer and Bruch’s membrane layer and \emph{en-face} OCTA images generated by maximum projection between internal limiting membrane layer and outer plexiform layer. We manually select and form a subset of OCTA-500 (191 from 3M and 254 from 6M), eliminating samples with severe quality issues on either the OCT domain or the OCTA domain. To be noted, the OCT and OCTA images in our subset are unpaired. We resize each image to $384 \times 384$. In this subset, 50 images are used for testing and the others are used for training.

\vspace{-0.1cm}
\subsection{Experimental Setting}
\label{ssec:setting}
\vspace{-0.1cm}
We train and evaluate our DCDA pipeline on unpaired UDA tasks in terms of both OCT to OCTA and OCTA to OCT. Such bi-directional experiments can effectively demonstrate the robustness of our framework. All compared methods and our DCDA are implemented with Pytorch using NVIDIA TITAN RTX GPUs. We use the Adam optimizer with a learning rate of $1 \times 10^{-4}$ and a weight decay of $1 \times 10^{-4}$ for all subnetworks included in DRST but no learning rate policy for the segmentation models. We first separately train DRST for 200 epochs and $\mathcal{F^{S}}$ for 100 epochs. Then we train the whole framework for another 800 epochs. We use Dice loss for $\mathcal{L}_{seg}$ and cross entropy loss for $\mathcal{L}_{ccl}$. The total loss is the summation of $\mathcal{L}_{seg}$ and $\mathcal{L}_{ccl}$. During testing, target domain images are directly inputted to $\mathcal{F^{T}}$ to get the corresponding predictions.

\vspace{-0.1cm}
\subsection{Results}
\label{ssec:result}
\vspace{-0.1cm}
All methods are evaluated using two metrics, i.e., Dice[\%] and 95\% Hausdorff Distance (HD[px]), the results of which are tabulated in Table \ref{tab:result1}. We compare DCDA with several SOTA UDA models including ADDA \cite{tzeng2017adversarial}, ADVENT \cite{vu2019advent}, AdaptSegNet \cite{tsai2018learning}, and CyCADA \cite{hoffman2018cycada}. Apparently, our proposed DCDA achieves the best RV segmentation performance on both OCT to OCTA and OCTA to OCT domain adaptation tasks among all methods. Most compared methods fail in a scenario of such a large domain gap. Intuitively, inverting either OCTA or OCT can reduce their domain gap (see Figure \ref{intro}). Therefore, we also conduct experiments by inverting the input images of the target domain. Related results are also shown in Table \ref{tab:result1}. It is obvious that in all directions and all experimental settings, our method achieves Dice scores that are closest to target-trained oracle, significantly outperforming other compared methods.

\begin{table*}[ht]
\centering
\small
\vspace{-0.5cm}
\caption{Quantitative evaluations of UDA methods for RV segmentation. We report UDA results both from OCT to OCTA and from OCTA to OCT. * denotes that the input images are inverted. Please note we do not invert input images for Oracle. A two tailed Student's t-test between DCDA and CyCADA identifies p-values that are much smaller than 0.05, indicating statistical significance.}
\vspace{0.2cm}
\label{tab:result1}
\begin{tabular}{c|ll|ll|ll|ll}
\hline
 Directions & \multicolumn{2}{c|}{OCT to OCTA} & \multicolumn{2}{c|}{OCTA to OCT} & \multicolumn{2}{c|}{OCT to OCTA *}    & \multicolumn{2}{c}{OCTA to OCT *}             \\ \hline
 Metrics                                & \multicolumn{1}{c}{Dice$\uparrow$} & \multicolumn{1}{c|}{HD$\downarrow$} & \multicolumn{1}{c}{Dice$\uparrow$} & \multicolumn{1}{c|}{HD$\downarrow$} & \multicolumn{1}{c}{Dice$\uparrow$} & \multicolumn{1}{c|}{HD$\downarrow$} & \multicolumn{1}{c}{Dice$\uparrow$} & \multicolumn{1}{c}{HD$\downarrow$} \\ \hline
 \multicolumn{1}{l|}{Source Only} &5.16±1.01                                                     &16.29±2.96                                               &5.71±2.04                                                    &28.65±12.35                           & 79.37±3.71
& 16.51±7.12
& 61.39±15.23
& 23.99±5.71                  \\
ADDA                             & 4.40±1.66                         & 13.72±5.98                       & 4.34±2.47                         & 48.29±31.08            & 80.14±3.13
& 14.48±8.02
& 58.17±16.47
& 29.38±8.58       \\
AdaptSegNet                      & 6.32±2.49                         & 21.16±4.06                   & 2.65±1.54                        & 34.59±16.41 & 74.81±3.70
& 14.89±7.96
& 58.58±10.73
& 32.58±6.94 \\
ADVENT                           & 26.22±4.23                         & 15.61±5.47            & 4.87±2.02                       & 35.69±13.18           & 78.82±2.90
& 12.96±5.70
& 6.11±1.84
& 45.06±9.03              \\
CyCADA                           & 60.21±5.28                          & 16.52±5.21                 & 18.38±2.67                         & 17.32±4.54          & 74.94±5.01
& 18.64±6.94
& \textbf{79.93±5.52}
& \textbf{10.71±4.55}        \\
DCDA (ours)                             & \textbf{83.85±3.51}                         & \textbf{9.13±4.49}                       &\textbf{79.28±6.16}                          & \textbf{15.97±6.08}   & \textbf{84.85±2.47}
 & \textbf{9.41±3.98}
 & 78.92±6.35 
& 15.97±7.46                          \\
Oracle                      & 87.48±1.97                        & 6.08±3.73                                   & 81.06±5.17                                        & 11.94±5.42 & 87.48±1.97                        & 6.08±3.73                                   & 81.06±5.17                                        & 11.94±5.42                              \\ \hline
\end{tabular}
\vspace{0.9cm}
\end{table*}

\begin{figure*}[htb]
	\centering
	\vspace{-1.0cm}
 	\setlength{\abovecaptionskip}{-0.2cm} 
 	\setlength{\belowcaptionskip}{-1cm}   
	\centerline{\includegraphics[height=8.3cm]{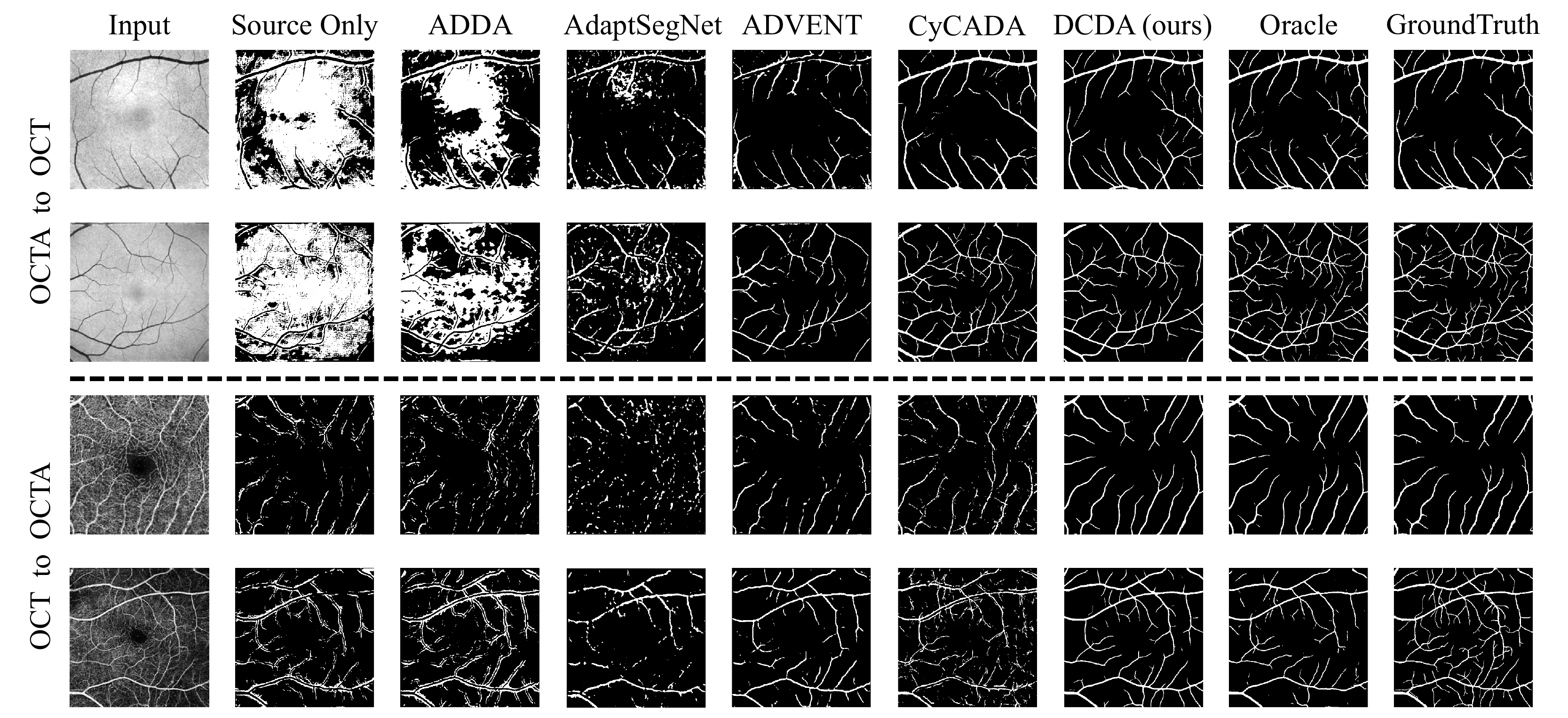}}
	\vspace{-0.3cm}
	\caption{Representative visualization results of RV segmentation in two domain adaptation directions. The inputs of the first and the third rows are 3M images and the inputs of the other two rows are 6M images.
	}\medskip
	\vspace{-0.3cm}
	\label{result_figure}
\end{figure*}

Representative visualization results are illustrated in Figure \ref{result_figure}. It is evident that our framework achieves superior performance on both 3M and 6M images and in both domain adaptation directions.

To evaluate the effectiveness of several key components in DCDA, we conduct ablation studies in Table \ref{tab:ablation}. We compare with our proposed framework without CCL (w/o $\mathcal{F^{S}}$), without the supervision of ground truth after the starting flag $\tau$ (w/o $\mathcal{L}_{seg}$) and without minimizing the distribution differences between the corresponding outputs of $\mathcal{F^{S}}$ and $\mathcal{F^{T}}$ (w/o $\mathcal{L}_{ccl}$). The results identify the significance of each component in our DCDA framework.

\vspace{-0.3cm}
\section{Conclusion}
\label{sec:conclusion}
\vspace{-0.2cm}
In this paper, we proposed and validated a novel framework for unsupervised domain adaptation. We disentangled images into content components and style codes and performed style transfer and image reconstruction. Adversarial training and cycle consistency were applied to encourage generators to produce style transferred images as real as possible.

Building on top of this, we further utilized a collaborative consistency learning module to successfully address UDA problems caused by large distribution differences between source domain and target domain. Through extensive quantitative and qualitative experiments, our proposed DCDA method was found to be much more superior than representative SOTA UDA methods, in terms of RV segmentation from OCT and OCTA images.

\begin{table}[htbp]
\centering
\large
\vspace{-0.7cm}
\caption{Ablation studies of our proposed framework.}
\vspace{0.2cm}
\label{tab:ablation}
\resizebox{8.1cm}{!}{
\begin{tabular}{c|cc|cc}
\hline
 Directions & \multicolumn{2}{c|}{OCT to OCTA} & \multicolumn{2}{c}{OCTA to OCT} \\ \hline
Metrics                                                   & \multicolumn{1}{c}{Dice$\uparrow$}  & HD$\downarrow$  & \multicolumn{1}{c}{Dice$\uparrow$}  & HD$\downarrow$  \\ \hline
Source Only  &5.16 &16.29 &5.71 &28.65     \\
w/o $\mathcal{F^{S}}$ & 55.71 & 51.39 & 66.68 & 27.03    \\
w/o $\mathcal{L}_{seg}$ & 81.28 & 13.03 & 76.99 & 17.35    \\
w/o $\mathcal{L}_{ccl}$ & 83.16 & 11.12 & 79.27 & 16.32    \\
Ours  & \textbf{83.85}  & \textbf{9.13} &\textbf{79.28} & \textbf{15.97}  \\
Oracle  & 87.48 & 6.08 & 81.06 & 11.94 \\ \hline
\end{tabular}}
\end{table}

\vspace{-0.3cm}
\section{Acknowledgments}
\vspace{-0.3cm}
This study was supported by the Shenzhen Basic Research Program (JCYJ20190809120205578); the National Natural Science Foundation of China (62071210); the Shenzhen Basic Research Program (JCYJ20200925153847004); the High-level University Fund (G02236002).



\small
\bibliographystyle{IEEEbib}
\bibliography{refs}

\end{document}